\documentclass[aps,prb,twocolumn,superscriptaddress,longbibliography]{revtex4-2}

\usepackage{hyperref} 
\usepackage{graphicx}
\usepackage{bm}
\usepackage{amsmath}
\usepackage{amssymb}
\usepackage{braket}
\usepackage[colorinlistoftodos]{todonotes}
\usepackage{xcolor}
\begin{document}
\title{Topological linear magnetoresistivity and thermoconductivity induced by noncentrosymmetric Berry curvature}

\author{ Min-Xue Yang}
\thanks{The authors contribute equally.}
\affiliation{National Laboratory of Solid State Microstructures,  School of Physics,
and Collaborative Innovation Center of Advanced Microstructures, Nanjing University, Nanjing 210093, China}

\author{Hai-Dong Li }
\thanks{The authors contribute equally.}
\affiliation{National Laboratory of Solid State Microstructures,  School of Physics,
and Collaborative Innovation Center of Advanced Microstructures, Nanjing University, Nanjing 210093, China}

\author{Wei Luo }
\affiliation{National Laboratory of Solid State Microstructures,  School of Physics,
and Collaborative Innovation Center of Advanced Microstructures, Nanjing University, Nanjing 210093, China}
\affiliation{School of Science, Jiangxi University of Science and Technology, Ganzhou 341000, China}

\author{Bingfeng Miao}
\affiliation{National Laboratory of Solid State Microstructures,  School of Physics,
and Collaborative Innovation Center of Advanced Microstructures, Nanjing University, Nanjing 210093, China}

\author{Wei Chen }
\email{Corresponding author: pchenweis@gmail.com}
\affiliation{National Laboratory of Solid State Microstructures,  School of Physics,
and Collaborative Innovation Center of Advanced Microstructures, Nanjing University, Nanjing 210093, China}

\author{D. Y. Xing }
\affiliation{National Laboratory of Solid State Microstructures,  School of Physics,
and Collaborative Innovation Center of Advanced Microstructures, Nanjing University, Nanjing 210093, China}

\begin{abstract}
The Berry curvature plays a key role in the magnetic transport of topological
materials. Yet, it is not clear whether the Berry 
curvature by itself can give rise to universal transport
phenomena with specific scaling behaviors. 
In this work, based on the semiclassical Boltzmann formalism and the symmetry analysis, we
show that the noncentrosymmetric distribution of the Berry curvature generally results in linear
magnetoresistivity and thermoconductivity both exhibiting the $B$-scaling behavior. We then study
such kind of topological linear magnetoresistivity in the 2D M$\text{n}$B$\text{i}$$_2$T$\text{e}_{4}$ flakes
and the 3D spin-orbit-coupled electron gas, the former showing good agreement with the experimental observations.
The difference between our mechanism and the conventional anisotropic magnetoresistance is elucidated.
Our theory proposes a universal scenario for the topological linear magnetoresistivity and thermoconductivity
and predicts such effects to occur in various materials, which also 
provides a reasonable explanation for the recent observations of linear magnetoresistivity.
\end{abstract}
\maketitle

\section{Introduction}\label{Intro}

The magnetic field effect on electron dynamics is an important manifestation
of the electron properties, which can be probed by the transport measurements of the magnetoresistivity (MR)~\cite{pippard1989magnetoresistance}. 
There are a variety of physical scenarios for the MR, which can usually be
discriminated by their specific scaling behaviors in the magnetic field $B$. For example,
the conventional positive magnetoresistance in metal possesses a $B^2$-scaling~\cite{ziman1972principles}
and the weak localization effect results in the dimension-dependent scaling behaviors of the MR with the
$\ln B$-scaling for 2D diffusion and $\sqrt[]{B}$-scaling for 3D diffusion, respectively~\cite{lee85rmp}.
In the past two decades, the study on topological materials~\cite{hasan2010rmp,qi2011rmp,lv2021rmp}
shows that the band topology introduces new ingredients to the MR scenarios. For example, 
the chiral anomaly and Berry curvature effect 
can induce negative MR with $B^2$-scaling in Weyl semimetals~\cite{son2013prb,burkov2014prl}
and the nontrivial Berry phase results in a sign change of the MR in topological insulators
and topological semimetals~\cite{lu11prl,garate12prb,lu14prl,lu2015prb,chen2019prl}, etc.

Different from the scaling laws mentioned above, there also exist
several scenarios which can give rise to linear MR (LMR)~\cite{gl2001prl,zyuzin2021prb,
xiao2020prb,nagaosa2016prl,ma2019prb,das19prb,zhang22arxiv,lu2015prb,chang2013science,deng2020science,jun2015prb,
liang2015nm,novak2015prb,Narayanan2015prl,zhao2015prx,tang2011acs,wang2012prb,
wang2012prl,Abrikosov98prb,feng2015prb,checkelsky2012np,gilbert2015nc}.
 In particular, there are two
types of LMR in the literature which are proportional to $B$ and $|B|$, respectively.
In the former case, the MR changes its sign as $B$ is inverted and so
exhibits nonreciprocity while in the latter case, the sign of the MR remains the same
despite the inversion of $B$.
Away from the quantum limit~\cite{Abrikosov98prb,chang2013science,lu2015prb,zhao2015prx,wang2012prl},
the LMR with $B$-scaling may originate from
the chiral scattering and self-field effect~\cite{gl2001prl}, chiral anomaly~\cite{nagaosa2016prl},
spin-orbit coupling and ferromagnetic momentum-dependent exchange interaction~\cite{zyuzin2021prb}, intravalley-scattering effects~\cite{xiao2020prb},
and complex spin configurations~\cite{gilbert2015nc}, etc;
Moreover, the MR with $|B|$-dependence is also observed in several experiments~\cite{liang2015nm,feng2015prb,parish2003nature},
which are attributed to the disorder-induced mobility fluctuation~\cite{parish2003nature,Narayanan2015prl}
and the shift of the Fermi surface~\cite{liang2015nm,feng2015prb}.
As one can see, different from other types of scaling laws
which are pertinent to certain mechanisms, the LMR can arise due to
a variety of physical origins. On the one hand, the phenomena associated with the LMR
indicate rich underlying physics; On the other hand, it becomes extremely challenging
to discriminate different scenarios from each other in the experiments.

\begin{figure}
    \centering
    \includegraphics[width=0.5\textwidth]{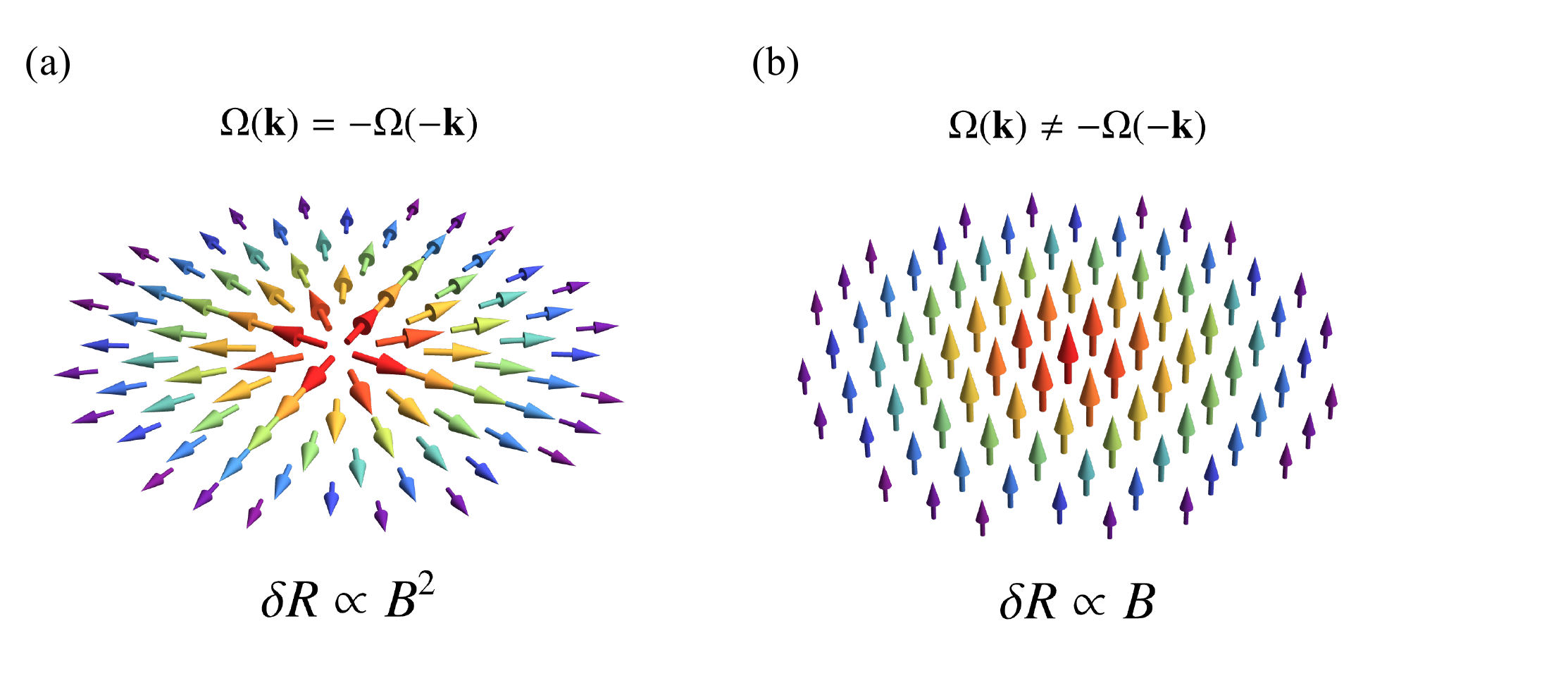}
    \caption{(Color online). Sketch of the (a) centrosymmetric
      and (b) noncentrosymmetric distributions 
    of the Berry curvature, which give rise to the magnetoresistivity
    with $B^2$-scaling and $B$-scaling, respectively.}
    \label{fig1}
    \end{figure}

In this work, we uncover a new and general mechanism for the nonreciprocal LMR with $B$-scaling,
dubbed \emph{topological LMR} (TLMR). It arises due to the noncentrosymmetric distribution of the Berry curvature
of the energy bands in the reciprocal space; see Fig.~\ref{fig1}.
In the semiclassical transport regime, the effects due to the Berry curvature mainly originate from
two aspects, the anomalous velocities~\cite{xiao2010rmp} and the correction to the density of 
states (DOS)~\cite{xiao2005prl,duval2006plb}.
It gives rise to several exotic magneto-transport
phenomena such as the anomalous Hall effect~\cite{nagaosa2010rmp,xiao2010rmp},
the negative MR in Weyl semimetals and topological insulators~\cite{son2013prb,burkov2014prl,dai17prl},
and the chiral magnetic effect~\cite{armitage2018rmp,chang2015prb,stephanov2012prl}, etc.
It was also discovered in some specific systems that the Berry curvature effect
can lead to nonreciprocal LMR~\cite{ma2019prb,das19prb,zhang22arxiv}. Yet, whether there exists
a general mechanism for the TLMR induced by 
the Berry curvature remains an open question. 

In this work, based on the semiclassical Boltzmann formalism and symmetry
analysis, we show that the noncentrosymmetric distribution of the Berry curvature in the reciprocal space
can generally give rise to the TLMR and similar topological linear thermoconductivity
(TLTC) with $B$-scaling in both 2D and 3D systems. 
We exemplify the general mechanism with two physical systems: the 2D flakes of topological material $\rm{MnBi}_2\rm{Te}_4$~\cite{deng2020science} 
and 3D electron gas with the Rashba or Dresselhaus spin-orbit coupling and a magnetic order~\cite{shen2004prb}. 
For the former case, the results calculated by our theory
are in good agreement with the observations in the transport experiment on the $\rm{MnBi}_2\rm{Te}_4$ flakes~\cite{deng2020science};
Meanwhile, we predict the TLMR effect in the 3D electron gas with spin-orbit coupling and leave it
to be explored in future experiments. Given that the main ingredient, noncentrosymmetric
distributed Berry curvature is ubiquitous,
the TLMR and TLTC effects are expected to exist 
in various systems with nonvanishing Berry curvature.

The rest of the paper is organized as follows.
The semiclassical transport theory is briefly introduced in Sec.~\ref{semi}.
In Sec.~\ref{symm}, a symmetry analysis on the occurrence of the TLMR
is given. Then the TLMR in the topological material
$\rm{MnBi}_2\rm{Te}_4$ is studied and 
the results are compared with the experimental observations in Sec.~\ref{mbt}.
In Sec.~\ref{rash}, we predict the TLMR to arise in the 3D spin-orbit coupled electron gas with a magnetic order.
In Sec.~\ref{md}, we discuss the phenomena associated with the multi-domain
structures and compare our scenario with that of the conventional anisotropic magnetoresistance.
The effect of the TLTC is discussed in Sec.~\ref{tc}.
Finally, some concluding remarks are given in Sec.~\ref{con}.


\section{Semiclassical transport Theory}\label{semi}
We consider the magnetic transport in the diffusive regime and describe
it by the semiclassical Boltzmann formalism. For a weak magnetic field
such that the electron cannot achieve the entire cyclotron motion 
along the closed orbit before getting scattered, the effect due to 
the Landau levels can be neglected. The semiclassical equations of 
motion for the electron in the energy bands with nonvanishing Berry curvature 
read~\cite{xiao2010rmp,sundaram1999prb}
\begin{equation}\label{1}
\dot{\bm{r}}=\frac{1}{\hbar} \nabla_{\bm{k}}\epsilon_{\bm{k}}-\dot{\bm{k}}\times \bm{\Omega_k},\qquad
\dot{\bm{k}}=-\frac{e}{\hbar} (\bm{E}+\dot{\bm{r}}\times \bm{{B}}),
\end{equation}
where $\bm{r}$ and $\bm{k}$ are the center position
and momentum of the wave packet, respectively,
$\bm{\Omega_k}=i \langle \nabla_{\bm{k}} u_{\bm{k}}|\times | \nabla_{\bm{k}} u_{\bm{k}}\rangle$
is the Berry curvature defined by the wave functions $|u_{\bm{k}}\rangle$, $\epsilon_{\bm{k}} $ is the energy,
$\bm{E}$ ($\bm{B}$) are the external electric (magnetic) field and $\hbar$ is the reduced Planck constant.

In the linear response regime, the effective velocity can be solved as (taking  $\bm{E}=0$)
\begin{equation}\label{2}
\dot{\bm{{r}}}=[\bm{v_k}+(e/\hbar) \bm{B} (\bm{v_k} \cdot{\bm{\Omega_k}})]/D_{\bm{k}},
\end{equation}
where $\bm{v_k}=(1/\hbar) \nabla_{\bm{k}} \epsilon_{\bm{k}}$ is the group velocity
and $D_{\bm{k}}$
is the correction to the DOS~\cite{xiao2005prl,duval2006plb} with
\begin{equation} \label{3}
D_{\bm{k}}=1+\frac{e}{\hbar} \bm{B} \cdot \bm{\Omega_{{k}}}.
\end{equation}
Similarly, the time derivative of the momentum is
\begin{equation}\label{4}
\dot{\bm{k}}=-\frac{e}{\hbar D_{\bm{k}}}[\bm{E}+\bm{v}_{\bm{k}}\times \bm{B}+\frac{e}{\hbar} (\bm{B}\cdot\bm{E})\bm{\Omega_k}].
\end{equation}

We employ the relaxation time approximation 
and make the assumption that various scattering processes are independent.
This enables us to employ the Mattiessen's rule $1/\tau=\sum_{i=1}^3 1/\tau_{i}$~\cite{ashcroft1976solid},
in which three main scenarios that dominate the transport are the scattering due to the
impurity, electron-electron interaction and phonon, characterized by
the relaxation time $\tau_{1}$, $\tau_{2}$ and $\tau_{3}$, respectively.  
Under the uniform and steady-state condition,
the semiclassical Boltzmann equation reduces to
\begin{equation}\label{5}
\dot{\bm{k}}\cdot \nabla_{\bm{k}}f=-\frac{f-f_0}{\tau},
\end{equation}
where $f$ is the actual distribution function of electrons under the action
of the external fields and
$f_0=1/[e^{(\epsilon_{\bm{k}}-\epsilon_F)/k_B T}+1]$
is the Fermi-Dirac function in equilibrium with a Fermi energy $\epsilon_F$. By combining
Eq.~\eqref{4} and Eq.~\eqref{5}, the distribution function 
deviated from the equilibrium to the first-order of $\bm{E}$ can be obtained as 
\begin{equation}\label{6}
f_1=\frac{e\tau}{D_{\bm{k}}} [\bm{E}+\frac{e}{\hbar} (\bm{E}\cdot{\bm{B}})\bm{\Omega_{k}}] \cdot\bm{v_k} \frac{\partial f_0}{\partial \epsilon_{\bm{k}}}.
\end{equation}
It gives rise to a finite current density with $D_{\bm{k}}$ is the 
correction to the DOS
\begin{equation}\label{7}
\bm{J}=-e \int \frac{d^3\bm{k}}{(2\pi)^3} f_1 D_{\bm{k}} \dot{\bm{r}}.
\end{equation}
By inserting Eqs.~\eqref{2} and \eqref{6} into the current expression, 
the longitudinal conductivity $\sigma_{\mu\mu}$ defined by $J_\mu=\sigma_{\mu\mu}E_\mu$ $(\mu=x,y,z)$
can be obtained as
\begin{equation}\label{8}
\sigma_{\mu\mu}=-\frac{e^2 \tau}{(2 \pi)^3} \int \frac{d^3 \bm{k}}{D_{\bm{k}}} (v^{\mu}_{\bm{k}}+\frac{e}{\hbar} B^\mu \bm{v_k}\cdot\bm{\Omega_k})^2
\frac{\partial f_0}{\partial \epsilon_{\bm{k}}}.
\end{equation}

\section{Symmetry Analysis on the TLMR}\label{symm}
The general condition for the occurrence of the TLMR (as well as the TLTC to be discussed in Sec.~\ref{tc}) can already be 
drawn from Eq.~\eqref{8} through the symmetry analysis
without going into details about the systems, which reflects the universality of the present scenario of
the TLMR. To see this,
we expand Eq.~\eqref{8} to the first order in $\bm{B}$ as
\begin{equation}\label{10}
\sigma^{(1)}_{\mu\mu}=\frac{e^3\tau }{(2\pi)^3\hbar}
\int d^3\bm{k}v^{\mu}_{\bm{k}}
[v_{\bm{k}}^{\mu}(\bm{B}\cdot{\bm{\Omega_k}})-2B^{\mu}(\bm{v_k}\cdot \bm{\Omega_k})]
\frac{\partial f_0}{\partial \epsilon_{\bm{k}}},
\end{equation}
where the two terms inside the square parenthesis come from the DOS correction $D_{\bm{k}}$ in Eq.~\eqref{3} and 
the anomalous velocity in Eq.~\eqref{2}, respectively. The expression contains the integral over 
the Berry curvature $\bm{\Omega}_{\bm{k}}$ around the Fermi surface, which can be 
regarded as a fictitious magnetic field
in the reciprocal space. The TLMR takes place
as $\sigma^{(1)}_{\mu\mu}$ possesses a finite value.

The symmetry restriction on the distribution of the Berry curvature
has a strong impact on $\sigma^{(1)}_{\mu\mu}$. Specifically,
in the presence of the time-reversal symmetry, the Berry curvature 
satisfies~\cite{xiao2010rmp}
\begin{equation}\label{9}
    \bm{\Omega_k}=-\bm{\Omega_{-k}},
    \end{equation} 
meaning a centrosymmetric distribution of the vector field 
in the momentum space as shown in Fig.~\ref{fig1}(a). Note that
the inversion operation reverses both the wave vector $\bm{k}$ and the vector field $\bm{\Omega}_{\bm{k}}$.
Then the integrals of both two terms in Eq.~\eqref{8} vanish
given that the velocity also obeys $\bm{v_k}=-\bm{v_{-k}}$ due to the time reversal symmetry
which stems from $\epsilon_{\bm{k}}=\epsilon_{-\bm{k}}$. In this case,
the magnetoconductivity (as well as the MR) to the lowest-order in $B$ is $\propto B^2$ meaning the absence
of the TLMR. 
Such negative quadratic MR has been proposed theoretically~\cite{son2013prb,burkov2014prl} and
confirmed experimentally~\cite{huang2015prx,ong2021nrp} in Weyl semimetals,
in which the centrosymmetric distribution of $\bm{\Omega}_{\bm{k}}$ sketched in Fig.~\ref{fig1}(a) is satisfied
for each Weyl cone although the whole system may not have the time-reversal symmetry.
This case can be understood as a time-reversal-like antiunitary symmetry defined locally
at each Weyl cone.

From the symmetry analysis above, one can see that the 
noncentrosymmetric distribution of the Berry curvature like that in Fig.~\ref{fig1}(b),
and the underlying symmetry restriction on the system
are essential for the TLMR. It should be noted that
mathematically, the noncentrosymmetric structures of both $\bm{\Omega}_{\bm{k}}$
and $\bm{v}_{\bm{k}}$ do not necessarily ensure the occurrence of the LMR
because the integral in Eq.~\eqref{10} exhibits a complex dependence
on both $\bm{\Omega}_{\bm{k}}$ and $\bm{v}_{\bm{k}}$. In reality, however,
it is reasonable to expect a finite $\sigma^{(1)}_{\mu\mu}$ in case that
the system has no symmetry restriction on $\bm{\Omega}_{\bm{k}}$ and $\bm{v}_{\bm{k}}$.

Furthermore, we show that the conclusion above maintains even
the effect due to the orbital magnetic moment is taken into account.
That is, in the system with time-reversal(-like) symmetry, 
neither Berry curvature nor orbit magnetic moment will induce TLMR.
The orbit magnetic moment originates from
the self-rotation of the wave packet~\cite{xiao2010rmp,chang1996prb}
and takes the form of 
\begin{equation}
    \bm{\mathcal{M}}_{\bm{k}}=-i \frac{e}{2 \hbar} \langle \nabla_{\bm{k}} u_{\bm{k}}| \times[H(\bm{k})-\epsilon({\bm{k}})]
    |\nabla_{\bm{k}} u_{\bm{k}}\rangle.
\end{equation}
where $H(\bm{k})=e^{-i \bm{k}\cdot \bm{r}} H e^{i \bm{k} \cdot \bm{r}}$ is 
the Bloch Hamiltonian.
Its coupling to the external magnetic field $\bm{B}$
results in a Zeeman-like term and the energy
becomes $\tilde{\epsilon}_{\bm{k}}=\epsilon_{\bm{k}}-\bm{\mathcal{M}}_{\bm{k}} \cdot {\bm{B}}$,
which modifies the group velocity of the electron to $\bm{\tilde{v}_{k}}=\bm{v_{k}}-\delta\bm{v}_{\bm{k}}$ with
$\delta\bm{v}_{\bm{k}}=\nabla_{\bm{k}}(\bm{\mathcal{M}}_{\bm{k}}\cdot\bm{B})/\hbar$.
As a result, the longitudinal conductivity in Eq.~\eqref{8} changes into~\cite{pal2010prb,dai17prl,yang2021prb}
\begin{equation}\label{11}
\tilde{\sigma}_{\mu\mu}=-\int \frac{d^3 \bm{k}}{(2 \pi)^3} \frac{e^2 \tau}{D_{\bm{k}}} (\tilde{v}^{\mu}_{\bm{k}}+\frac{e}{\hbar} B^\mu \bm{\tilde{v}_k}\cdot\bm{\Omega_k})^2
\frac{\partial f_0}{\partial \tilde{\epsilon}_{\bm{k}}}.
\end{equation}
It can be proved that $\bm{\mathcal{M}}_{\bm{k}}$ possesses the same symmetry as that of the Berry curvature
if the system possesses the time-reversal(-like) symmetry, i.e.,
\begin{equation}
    \bm{\mathcal{M}}_{\bm{k}}=-\bm{\mathcal{M}}_{-\bm{k}},
\end{equation}
given that the former is induced by the latter. 
Thus, we have $\delta\bm{v_{k}}=\delta\bm{v_{-k}}$.
{{The effect due to the orbital magnetic moment contributes additional terms to Eq.~\eqref{10} as
\begin{equation}
    \delta\sigma^{(1)}_{\mu\mu}=\int\frac{d^3\bm{k}}{(2\pi)^3}e^2\tau [v^{\mu}_{\bm{k}}\delta v^{\mu}_{\bm{k}}\frac{\partial f_0}{\partial \epsilon_{\bm{k}}}
    +(v^{\mu}_{\bm{k}})^2 (\bm{\mathcal{M}_k}\cdot{\bm{B}}) \frac{\partial^2 f_0}{\partial \epsilon^2_{\bm{k}}}]
\end{equation}
However, such terms vanish as well due to the symmetry restriction 
$\bm{\mathcal{M}}_{\bm{k}}=-\bm{\mathcal{M}}_{-\bm{k}}, \delta\bm{v_{k}}=\delta\bm{v_{-k}}$ and $\bm{v_{k}}=-\bm{v_{-k}}$. }}Therefore,
the inclusion of the effect due to the orbital
magnetic moment does not change our conclusion, that is,
the noncentrosymmetric distribution of the Berry curvature
is key to the TLMR. 

The aforementioned scenario of the TLMR is quite general
and is expected to exist in a variety of materials
with nonvanishing Berry curvature distribution, no matter in 2D or 3D.
It does not
even require the whole Bloch bands to be topologically nontrivial, 
because only the Berry curvature of the states
near the Fermi energy dominates the effect.
 Here, we would like to point out
that although the noncentrosymmetric distribution of the Berry curvature provides
a general scenario for the TLMR, it is
not a necessary condition for it. From Eq.~\eqref{10},
the LMR can also occur under the condition $\bm{\Omega_k}=-\bm{\Omega_{-k}}$ 
and $\bm{v_k}\neq-\bm{v_{-k}}$, of which a typical example is the type II Weyl semimetal~\cite{das19prb,ma2019prb}.
Such a scenario is not the focus of this work.
For clarification, we compare different scenarios of the MR 
due to the Berry curvature effect in table~\ref{table}.

\begin{table*} 
    \centering
    \begin{tabular}{|c|c|c|c|}
        \hline
              $\bm{\Omega}_{\bm{k}}=-\bm{\Omega}_{-\bm{k}}$ & $\bm{v}_{\bm{k}}=-\bm{v}_{-\bm{k}}$ & $B$-Scaling of MR & Examples \\
              \hline
              $\times$ & $\times (\checkmark)$ &$B$  & MnBi$_2$Te$_4$ in Sec~\ref{mbt}, 
              electron gas in Sec~\ref{rash}, 
              magnetic Weyl semimetal~\cite{zhang22arxiv} \\
              \hline
              $\checkmark$ & $\times$ & $B$ & Type-II Weyl \cite{ma2019prb,das19prb} \\
               \hline
               $\checkmark$ & $\checkmark$ & $B^2$ & Type-I Weyl,$etc$ \cite{nagaosa2020nr} \\
               \hline
      \end{tabular}
      \caption{Symmetry conditions and the corresponding $B$-scaling of the MR. The last 
          column shows typical examples. \label{table}}
    \end{table*}

To show the universality of the noncentrosymmetric Berry curvature induced TLMR, 
in the following sections, we will study the TLMR in two
specific systems, the 2D MnBi$_2$Te$_4$ flakes
and the 3D spin-orbit coupled electron gas. The generalization of our theory
to other systems is straightforward.

\section{TLMR in 2D M$\text{n}$B$\text{i}$$_2$T$\text{e}_{4}$ flakes}\label{mbt}

We first consider the 2D material MnBi$_2$Te$_4$, which is the first
discovered intrinsic magnetic topological insulator that exhibits 
quantum anomalous Hall effect and has sparked
a surge of research interest~\cite{deng2020science,li2019sa,sass2020nano,fu2020sa}.
One unique feature of the 2D systems is that the Berry curvature possesses only the $z$-component
so that the MR originates solely from its coupling to the $B_z$ component of the magnetic field.
Because the electron transport takes place within
the 2D plane, the in-plane velocity and the out-of-plane Berry curvature
ensures $\bm{v}_{\bm{k}}\cdot\bm{\Omega}_{\bm{k}}=0$ in Eq.~\eqref{2}, \emph{i.e.},
the absence of the anomalous velocity.
Therefore, the MR stems entirely from the DOS correction $D_{\bm{k}}$ induced by the Berry curvature
according to Eq.~\eqref{8}.


The LMR with $B$-scaling has been observed 
in the transition region between the two opposite quantum anomalous Hall plateaus; see Fig.~2 of
Ref.~\cite{deng2020science}. In this region, the nonvanishing longitudinal resistance indicates
the participation of the bulk states in the transport. Out side of the
transition region, the Fermi energy should lie in the gap between
the conduction and valence bands.
We here employ the effective Hamiltonian of the $\rm{Mn}\rm{Bi}_2\rm{Te}_4$ flakes~\cite{deng2020science,wang2014prb}
\begin{equation} \label{12}
\begin{split}
H_{\text{MBT}}=
\left(
\begin{array}{cccc}
\Delta+m_{\bm{k}}&ivk_-&0&0\\
-ivk_+&-\Delta-m_{\bm{k}}&0&0\\
0&0&-\Delta+m_{\bm{k}}&-ivk_+\\
0&0&ivk_-&\Delta-m_{\bm{k}}\\
\end{array}
\right),\\
\end{split}
\end{equation}  
where $v$ is the velocity of the surface states, $k_{\pm}=k_x\pm ik_y$ is defined by the 2D wave vectors,
$m_{\bm{k}}=m_0+m_1 (k_x^2+k_y^2)$ captures the tunneling effect between the
top and bottom surface states with $m_0,m_1$ the relevant parameters,
and $\Delta$ is the exchange field along the $z$-direction due to
the ferromagnetic order. Whether the Dirac mass is inverted
or not at the $\Gamma$ point determines the
Chern number $\mathcal{C}$ of the valence bands and specifically, $\mathcal{C}=0$
for $|\Delta|<|m_0|$ and $\mathcal{C}=\mathrm{sgn}(\Delta)$ for 
$|\Delta|>|m_0|$, respectively~\cite{deng2020science}.
By diagonalizing the
Hamiltonian $H_{\text{MBT}}$, we obtain four bands with dispersions
$\epsilon_{u}^\pm=\pm\sqrt{(\Delta+m_{\bm{k}})^2+v^2k^2},
\epsilon_{l}^\pm=\pm\sqrt{(\Delta-m_{\bm{k}})^2+v^2k^2}$, where the subscripts $u, l$
denote the upper and lower blocks of Eq.~\eqref{12}, respectively.

\begin{figure}[t!]
    \centering
    \includegraphics[width=0.5\textwidth]{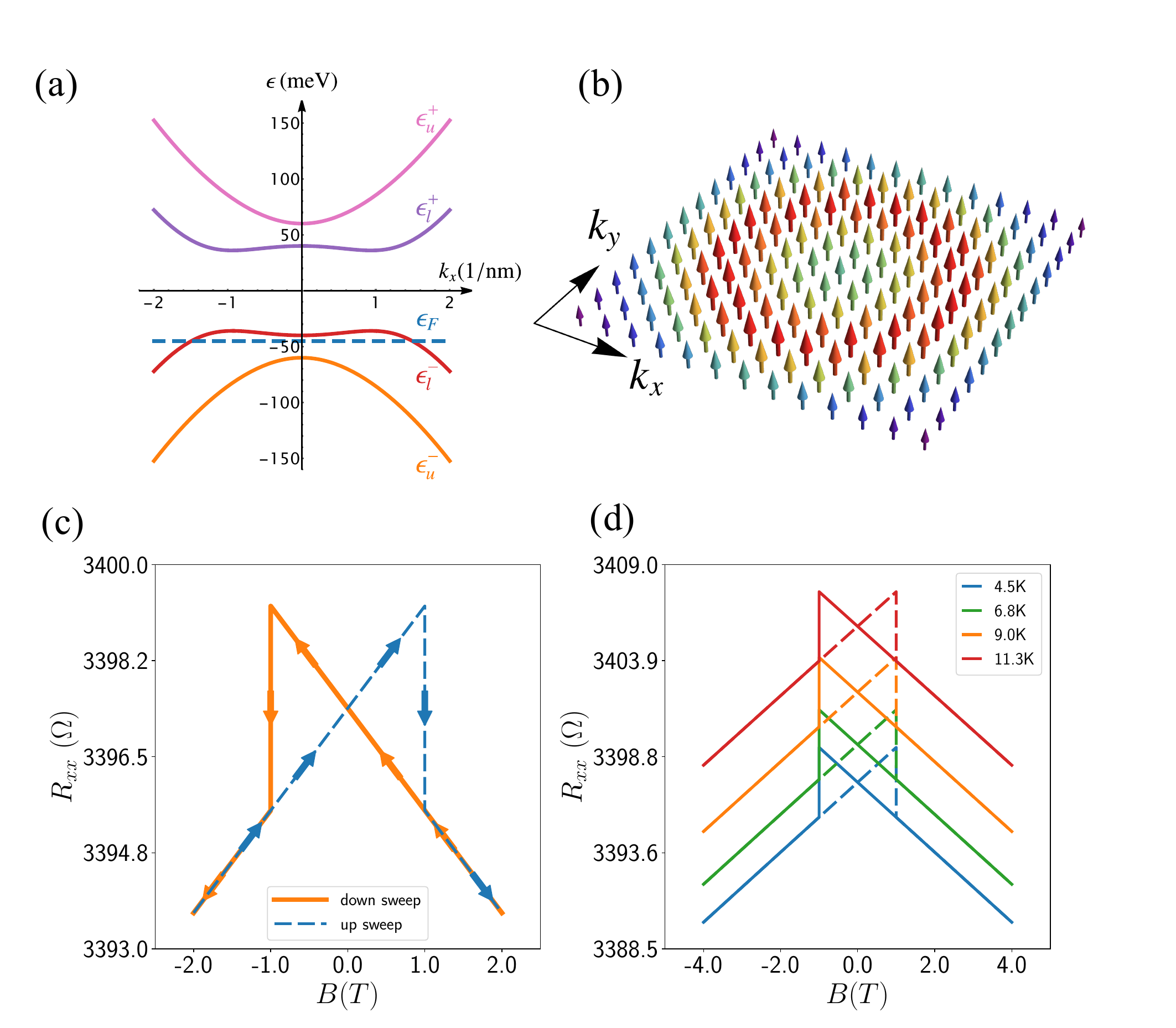}
    \caption{(Color online). (a)  Energy spectra of the surface
    state of $\rm{MnBi}_2\rm{Te}_4$ ($k_y=0$). 
    The dashed blue line represents the position of the Fermi energy.
    (b) The distribution of the Berry curvature of $\rm{MnBi}_2\rm{Te}_4$.
    (c) The TLMR as a function of the magnetic field at $T=1.2$ K. The arrowed dashed line  
    represents the up sweep and the arrowed solid line represents the down sweep. 
    (d) TLMR at different temperatures. The relevant
    parameters are $\tau=10^{-12}$ s, $\epsilon_F=-45$ meV, $|\Delta|=50$ meV,
    $m_0=10$ meV, $m_1=20$ meV\,$\rm{nm^2}$, $v=30$ meV\,nm.}
    \label{fig2}
    \end{figure}

To involve the bulk contribution to the longitudinal transport
as was observed in the experiment, we consider 
the Fermi energy $\epsilon_F$ intersects the upper valence band in the 
transition region as illustrated in Fig.~\ref{fig2}(a).
In particular, the upper valence band is referred to
$\epsilon_{l}^-$ and $\epsilon_{u}^-$ for $\Delta>0$ and $\Delta<0$, 
with the corresponding eigenstates {
$|v_l^-\rangle =[0,0,-i vk_+,-2\sqrt{(\Delta-m_{\bm{k}})^2+v^2 k^2} \sin^2{\theta_l}]^{T}/N_l$ and $|v_u^-\rangle=
[i v k_{-},-2\sqrt{(\Delta+m_{\bm{k}})^2+v^2 k^2} \cos^2{\theta_u},0,0]^T/N_u$, respectively,
whose normalization factors are $N_l=2\sqrt{(\Delta-m_{\bm{k}})^2+v^2 k^2} \sin{\theta_l}, 
N_u= 2\sqrt{(\Delta+m_{\bm{k}})^2+v^2 k^2} \cos{\theta_u}$ with the angles $\theta_{u,l} \in [0,\pi/2]$ defined by 
$\tan{2\theta_l}=v k/[\Delta-m_{\bm{k}}], 
\tan{2\theta_u}=v k/(\Delta+m_{\bm{k}})$}. The Berry curvature functions for the two bands are
\begin{equation}\label{13}
    \begin{split}
&\bm{\Omega}^{-}_{\bm{k}l}=-\frac{v^2 (m_0-m_1 k^2-\Delta)}{2[(\Delta-m_{\bm{k}})^2+v^2k^2]^\frac{3}{2}}\bm{e}_z,\\
&\bm{\Omega}^{-}_{\bm{k}u}=\frac{v^2 (m_0-m_1 k^2+\Delta)}{2[(\Delta+m_{\bm{k}})^2+v^2k^2]^\frac{3}{2}}\bm{e}_z.\\\
    \end{split}
\end{equation}
We plot the Berry curvature distribution $\bm{\Omega}_{\bm{k}l}^-$ for $\Delta>0$ in Fig.~\ref{fig2}(b). One
can see that $\bm{\Omega}_{\bm{k}}^-$ is even under the inversion of $\bm{k}$ so that it possesses a
noncentrosymmetric distribution.
Therefore, the MnBi$_2$Te$_4$ flakes should exhibit the TLMR according to the previous symmetry analysis.

The longitudinal conductivity is contributed by the electrons near the Fermi surface
so that we calculate $\sigma_{xx}$ in Eq.~\eqref{8} by inserting the parameters
$\bm{v}_{\bm{k}}$ and $\bm{\Omega}_{\bm{k}}$ with their values
in the relevant band, which is the top valence band here [cf. Fig.~\ref{fig2}(a)]. 
A magnetic field in the $z$ direction that is perpendicular to 
the flake is assumed. The resistivity is then obtained through the relation 
$R_{xx}=\sigma_{xx}/(\sigma^2_{xx}+\sigma^2_{xy})$ with
\begin{equation}
\sigma_{xy}=\frac{e^2}{{(2\pi)^2}\hbar}\sum_{i=u,l,\alpha=\pm}\int d^2 \bm{k}f_0(\bm{k})(\bm{\Omega}_{\bm{k}i}^{\alpha}\cdot \bm{e}_z)
\end{equation} 
being the Hall conductivity. The numerical results of $R_{xx}$ as a function of $B$ are plotted in 
Fig.~\ref{fig2}(c), in which the hysteresis effect in accordance with the experimental observation
has been taken into account.
The up-and-down sweeps of the magnetic field are marked by the arrowed
dashed and solid lines, respectively.
The coercive field is chosen to be $B_c=1$ T in the calculation,
close to the value in the experiment~\cite{deng2020science}. 
The results in Fig.~\ref{fig3}(c) clearly show the TLMR with $B$-scaling, as expected.
Here, both $\sigma_{xx}$ and $R_{xx}$ exhibit linear dependence on $B$,
because the change of $\sigma_{xx}$ by the magnetic field is small compared
with its initial value.

The hysteresis loops in Fig.~\ref{fig2}(c) can be understood as follows.
Starting with $B<-B_c$, and the ferromagnetic moment is aligned in the same
direction with the magnetic field with $\Delta>0$. As $B$ increases, the system exhibits a 
TLMR with a positive slope, \emph{i.e.}, $\Delta R_{xx}=R_{xx}(B)-R_{xx}(0)\propto B$ 
until $B$ reaches the coercive field $B_c$. In this region, 
$\Delta R_{xx}$ changes its sign as $B$ is inverted, showing nonreciprocity.
As $B$ exceeds $B_c$, $\Delta$ undergoes an abrupt flipping 
to lower the energy of the system, which results in a jump of the MR. Similar discontinuity
takes place for the Hall resistance as well, which changes from the quantized value
$h/e^2$ to $-h/e^2$. Both the abrupt changes of the longitudinal MR and the 
Hall resistance stem from the sign reversal of the Berry curvature.
The down sweep [arrowed solid line in Fig.~\ref{fig2}(c)] that starts
from $B>B_c$ and $\Delta>0$ can be understood in a similar way.
The TLMR feature in Fig.~\ref{fig2}(c) obtained by our theory
is in good agreement with the experimental observations~\cite{deng2020science}.

Furthermore, we plot the MR for different temperatures in Fig.~\ref{fig2}(d) [the
exchange field exhibits a very weak temperature dependence 
below the Néel temperature~\cite{deng2020sci,otrokov2019nature}] and has been chosen to be constant, 
which exhibit the same features and temperature dependence as those observed in
the experiment (see Fig.~2c in Ref.~\cite{deng2020science}).
As a result, the scenario of the TLMR proposed here
provides a reasonable and promising explanation 
for the MR observed in the $\rm{MnBi}_{2}{Te}_{4}$ flakes.

\section{TLMR in 3D spin-orbit-coupled electron gas}\label{rash}
The general scenario of the TLMR is not limited to the 2D systems. To illustrate its universality,
in the second example we investigate the TLMR in 
the 3D systems with spin-orbit coupling and a ferromagnetic order.
Consider first 3D model with the Rashba effect, 
which can be realized in BiTeI~\cite{bahramy2011prb,ishizaka2011nm,murakawa2013detection}.
The corresponding Hamiltonian is
\begin{equation} \label{14}
H_R=\frac{\hbar^2 k^2}{2 m}+\lambda (\bm{k}\times \bm{\sigma})\cdot \bm{e}_z+\Delta \sigma_z,
\end{equation}
where inversion symmetry is broken along the $z$-axis,
$m$ is the effective mass, the wave vector $\bm{k}=(k_x, k_y, k_z)$ 
consist of three components, $\lambda$ is the 
strength of the Rashba spin-orbit coupling, $\Delta$ is the Zeeman exchange field due
to the ferromagnetic order, and
$\bm{\sigma}=(\sigma_x,\sigma_y,\sigma_z)$ are the Pauli matrices for the spin. 
Without the exchange field the system possesses the time-reversal symmetry
and so cannot exhibit the TLMR.

\begin{figure}
    \centering
    \includegraphics[width=0.5\textwidth]{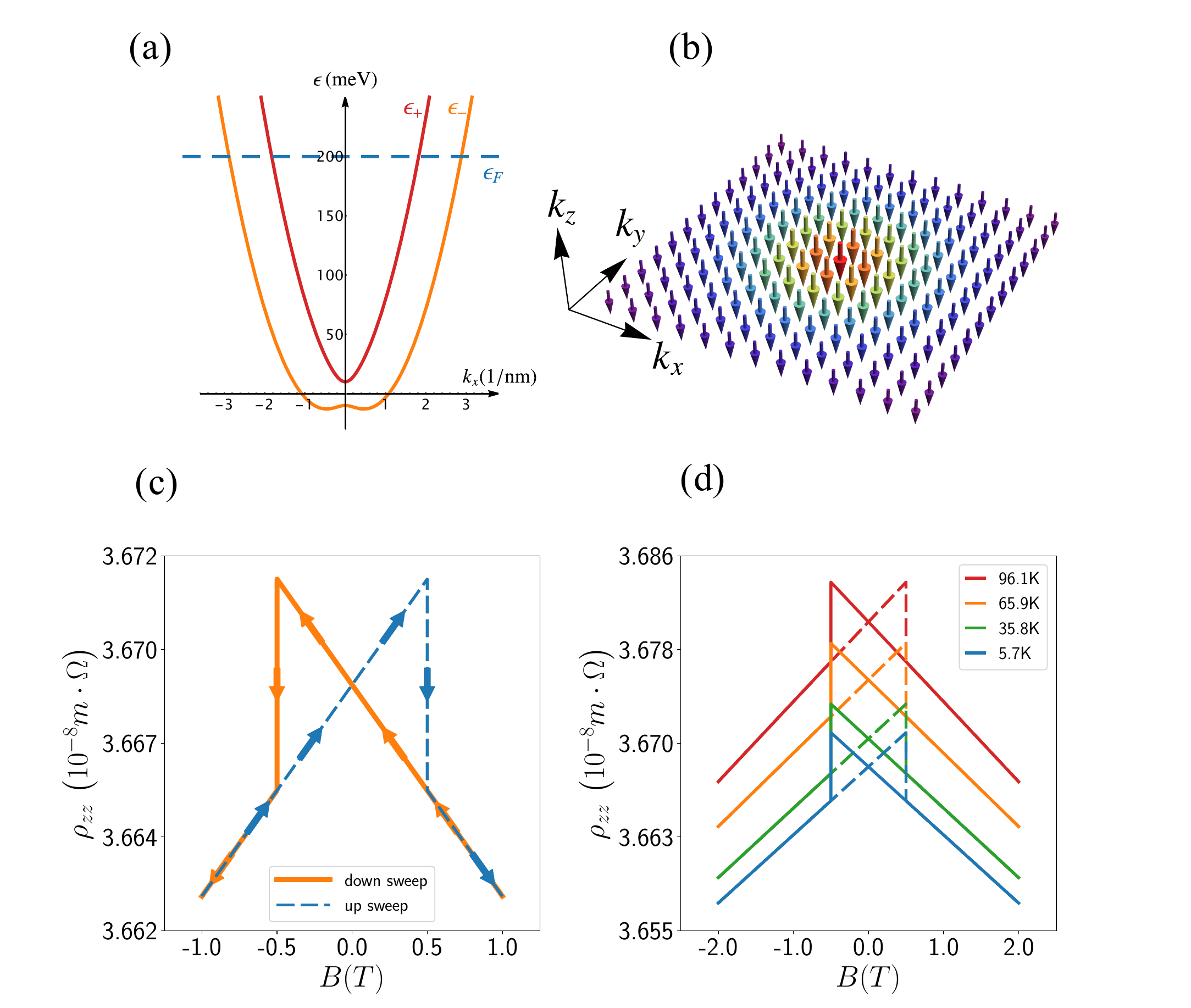}
    \caption{(Color online). (a) Energy spectra of the spin-orbit coupled
    electron gas ($k_z=0,k_y=0$). 
    The dashed blue line represents the position of the Fermi energy.
    (b) The distribution of the Berry curvature.
    (c) The TLMR as a function of the magnetic field at $T=4.7$ K. The arrowed dashed line  
    represents the up sweep and the arrowed solid line represents the down sweep. 
    (d) TLMR at different temperatures. The relevant
    parameters are $\tau=10^{-12}$ s, $\lambda=40$ meV, $\epsilon_F=200$ meV,$|\Delta|=10$ meV.}
    \label{fig3}
    \end{figure}

The energies of the two bands are $\epsilon_{\pm}=\hbar^2 k^2/2m\pm \sqrt{\lambda^2(k_x^2+k_y^2)+\Delta^2}$ 
as shown in Fig.~\ref{fig3}(a), whose corresponding wavefunctions are
{$|u_{\pm\bm{k}}\rangle=[\lambda(ik_x+k_y),\sqrt{\lambda^2 (k_x^2+k_y^2)+\Delta^2}(\cos{2\theta}\mp 1)]^{T}/N_{\pm}$ 
with the normalization factors
$N_{+}=2\sqrt{\Delta^2+\lambda^2 (k_x^2+k_y^2)} \sin{\theta}, N_{-}=2\sqrt{\Delta^2+\lambda^2 (k_x^2+k_y^2)} \cos{\theta}$ with
the angle $\theta \in [0,\pi/2]$ defined by
$\tan{2\theta}= \sqrt{\lambda^2 (k_x^2+k_y^2)}/\Delta$. }
The Berry curvature distribution of the two bands are
\begin{equation}\label{15}
\bm{\Omega}^{R\pm}_{\bm{k}}=\mp\frac{\Delta \lambda^2}{2[\Delta^2+(k_x^2+k_y^2)\lambda^2]^\frac{3}{2}} \bm{e}_z.
\end{equation}
Similar to those in Eq.~\eqref{13} for the 2D $\rm{Mn}\rm{Bi}_2\rm{Te}_4$, the Berry curvature here
contains only the $z$-component and possesses a noncentrosymmetric distribution as well;
see Fig.~\ref{fig3}(b) for the distribution of the Berry curvature 
in a specific $k_z$ plane. The Berry curvature does not rely on 
$k_z$ so that its noncentrosymmetric distribution in the 3D 
reciprocal space can be inferred from Fig.~\ref{fig3}(b).

We consider the Fermi energy $\epsilon_F$ that intersects
both bands as shown in Fig.~\ref{fig3}(a). Without loss of generality, 
the electric and magnetic fields
are both set to the $z$-direction.
The conductivity $\sigma_{zz}$ in Eq.~\eqref{8} is solved numerically 
taking into account the contributions from both bands. 
The 3D resistivity $\rho_{zz}=1/\sigma_{zz}$ is plotted in Fig.~\ref{fig3}(c), 
where we have assumed the existence of a single magnetic domain with a coercive field $B_c^R=0.5$ T.
The numerical results resemble those for the 2D MnBi$_2$Te$_4$ flakes,
again, consistent with the symmetry analysis in Sec.~\ref{symm}. 
The temperature dependence of the TLMR is shown in Fig.~\ref{fig3}(d),
which is also similar to the case of the $\rm{MnBi}_{2}{Te}_{4}$ [cf. Fig.~\ref{fig2}(d)].

Apart from the Rashba-type spin-orbit coupling, {the Dresselhaus-type spin-orbit coupling may also 
exist in 3D systems such as GaAs/AlGaAs quantum wells \cite{studer2010prb}
due to the breaking of the bulk inversion symmetry}~\cite{dresselhaus1955pr},
which can be described by
\begin{equation} \label{17}
H_D=\frac{\hbar^2 k^2}{2 m}+\beta (k_x \sigma_x-k_y \sigma_y) +\Delta \sigma_z,
\end{equation}
where $\beta$ is the strength of the Dresselhaus spin-orbit coupling. 
The two types of spin-orbit
coupling result in the same form of the energy spectra (with the replacement $\lambda\to \beta$) but different spin textures.
Accordingly, the Berry curvature for both bands takes the form of
\begin{equation} \label{18}
\bm{\Omega}^{D\pm}_{\bm{k}}=\pm \frac{ \Delta \beta^2}{2[\Delta^2+(k_x^2+k_y^2)\beta^2]^\frac{3}{2}} \bm{e}_z,
\end{equation}
which is similar to Eq.~\eqref{15} but with a sign
reversal for each band. Given that the sign of the TLMR 
is determined by the direction of the Berry curvature 
as indicated in Eq.~\eqref{10}, it can be expected that
the TLMR induced by the Dresselhaus spin-orbit coupling
possesses the opposite slope compared with that
due to the Rashba spin-orbit coupling.



\section{multi-domain structures and differences with AMR}\label{md}

\begin{figure}
    \centering
    \includegraphics[width=0.5\textwidth]{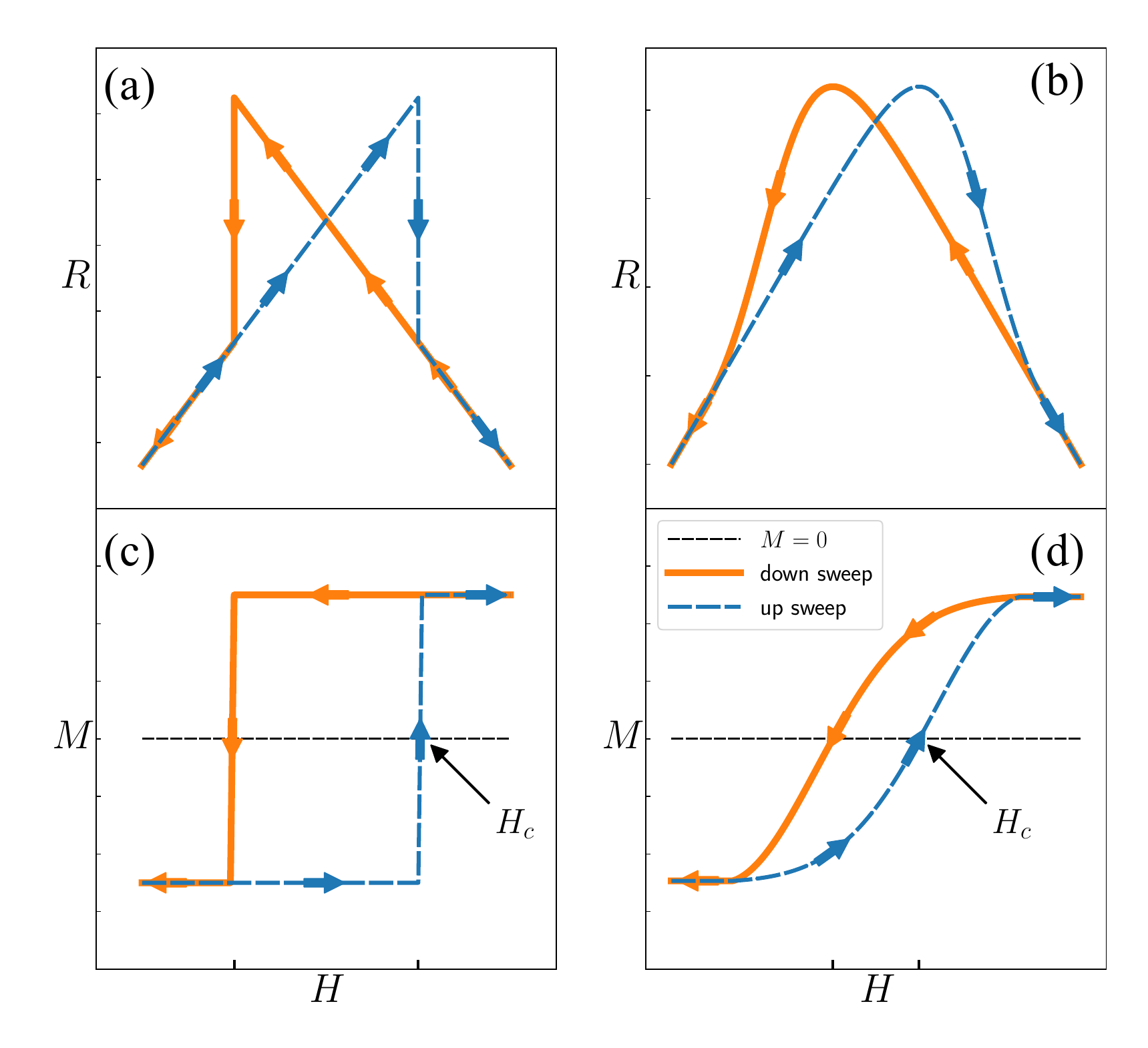}
    \caption{(Color online). (a) { Schematic diagram of} the TLMR in the single domain regime with
    the corresponding hysteresis loop of the magnetization in (c).
    (b)  Schematic diagram of} the TLMR in the multi-domain regime with
    the corresponding hysteresis loop of the magnetization in (d).
    \label{fig4}
    \end{figure}

In the examples discussed in the previous two sections, we have assumed that
the ferromagnetic material possesses a single magnetic domain, 
which results in the TLMR with the shapes sketched in Figs.~\ref{fig2}(c) and \ref{fig3}(c). 
Due to the magnetic structure of a single domain, the hysteresis loops of the magnetization
exhibits an abrupt flipping as shown in Fig.~\ref{fig4}(c),
which is responsible for the the discontinuity of the TLMR in Fig.~\ref{fig4}(a).
For the MnBi$_2$Te$_4$ flakes, it is exactly the case and has been verified in the experiment.
Meanwhile, in many ferromagnetic materials, multi-domain structures usually arise,
which give rise to the conventional hysteresis loop of the magnetization as sketched in Fig.~\ref{fig4}(d).
Recall that the TLMR here is induced by the Berry curvature configurations which is
determined by the (average) magnetization $M$ or equivalently, the Zeeman field $\Delta$ in previous discussions.
Therefore, it can be inferred that a continuous change of $M$ with the magnetic field $H$ leads to a smooth
variation of the Berry curvature as well. Accordingly, the MR is expected to change from the configuration
in Fig.~\ref{fig4}(a) to that in Fig.~\ref{fig4}(b), in which the abrupt
jump of the MR disappears. Nevertheless, the linear $B$ ($H$)-scaling persists
for a small magnetic field around zero.

It is worth noting that the 
LMR with $B$-scaling has also been 
reported in a variety of materials like Ge(111)~\cite{guillet2020prl,Guillet2021prb} and Co/Pd thin-film~\cite{gilbert2015nc},
the latter possessing the same MR configurations as those in Fig.~\ref{fig4}(b).
In these materials, the LMR was explained by the conventional anisotropic magnetoresistance (AMR) mechanism~\cite{mcguire1975ieee}.
Therefore, it is important to discriminate our scenario of the TLMR
in the multi-domain regime from that of the conventional AMR.
Physically, the main difference between the TLMR and the AMR is that
the former is induced by the joint action of the Berry curvature and the magnetic field
while the latter stems directly from the change of the magnetization rather than 
the magnetic field itself,
which possess different manifestations.
If the magnetization of the system remains 
unchanged as the magnetic field varies, which is just the case
in the regime of the single magnetic domain as shown in Fig.~\ref{fig4}(c)
or the saturation magnetization [cf. Fig.~5(b) in Ref.~\cite{gilbert2015nc} and Fig.~8 in Ref.~\cite{gil2005prb}],
the resistivity does not change for the AMR scenario.
In stark contrast, the TLMR shows up even if the magnetization remains constant in these regions. 
However, in the region where the magnetization changes smoothly with
the magnetic field, the MR may look similar for both the AMR and TLMR scenarios [compare Figs.~\ref{fig4}(b,d) with 
Fig.~5(b) in Ref.~\cite{gilbert2015nc} and Fig.~8 in Ref.~\cite{gil2005prb}].
In this case, further analysis of the results beyond such parametric regions is required
to discriminate the two scenarios.

\section{topological linear thermoconductivity}\label{tc}
\begin{figure}[t!]
    \centering
    \includegraphics[width=0.5\textwidth]{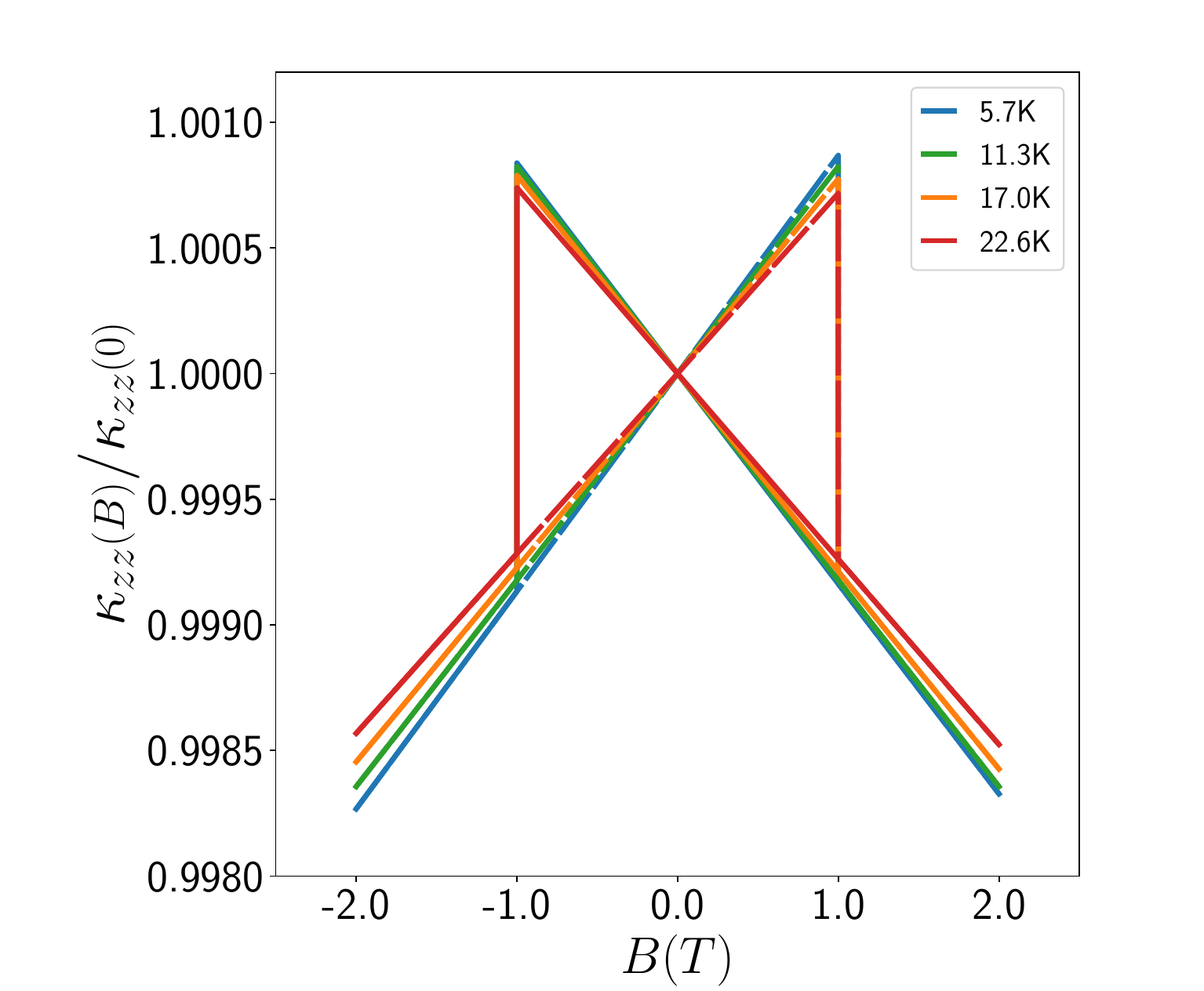}
    \caption{(Color online).
    TLTC at different temperatures. The dashed (solid) lines
    represent up (down) sweep.  The relevant
    parameters are $\tau=10^{-12}$ s, $\lambda=40$ meV, $\epsilon_F=-45$ meV, $|\Delta|=50$ meV,
    $m_0=10$ meV, $m_1=20$ meV $\rm{nm^2}$, $v=30$ meV nm.}
    \label{fig5}
    \end{figure}

So far, we have focused on the effect in the charge 
transport. Given that the resistivity is a specific type
of transport coefficient, 
the present scenario can be generalized straightforwardly to the other transport coefficients such as 
the thermoconductivity~\cite{xiao2006prl}. 
Similar to the TLMR, the TLTC
can also arise due to the same scenario.
Specifically, the $z$-direction thermoconductivity
$\kappa_{zz}$ is defined by~\cite{fiete2014prb,jiang2022prl,xiao2006prl}
\begin{equation}\label{19}
    \kappa_{zz}=L_{zz}^{22}-L_{zz}^{21} (L_{zz}^{11})^{-1} L_{zz}^{12},
\end{equation}
with $L_{zz}^{11}=\sigma_{zz}$ the longitudinal conductivity in Eq.~\eqref{8} and the other transport coefficients calculated by
\begin{equation}\label{20}
    \begin{split}
        L_{zz}^{21}&=TL_{zz}^{12}=\int \frac{d^3 \bm{k}}{(2 \pi)^3} \frac{e \tau}{D_{\bm{k}}} (v^{z}_{\bm{k}}+\frac{e}{\hbar} B^z \bm{v_k}\cdot\bm{\Omega_k})^2 (\epsilon_{\bm{k}}-\epsilon_F)
        \frac{\partial f_0}{\partial \epsilon_{\bm{k}}},\\
        L_{zz}^{22}&=-\int \frac{d^3 \bm{k}}{(2 \pi)^3} \frac{\tau}{TD_{\bm{k}}} (v^{z}_{\bm{k}}+\frac{e}{\hbar} B^z \bm{v_k}\cdot\bm{\Omega_k})^2 (\epsilon_{\bm{k}}-\epsilon_F)^2
        \frac{\partial f_0}{\partial \epsilon_{\bm{k}}}.
    \end{split}
\end{equation}
We here focus on
the elastic scattering in the heat transport which requires
a low temperature such that $\tau\simeq \tau_1$.
All these transport coefficients possess the similar forms of Eq.~\eqref{8}
and in particular, have the same $\bm{B}$-dependence.
Therefore, the symmetry analysis concludes that linear terms in $B$
arise for all the coefficients $L_{zz}^{11}, L_{zz}^{12}, L_{zz}^{21}$ and $L_{zz}^{22}$
as long as the distribution of the Berry curvature is noncentrosymmetric in the reciprocal space.
Furthermore, by noting that the linear terms in $B$ of all the transport coefficients
is much smaller than the constant terms, the thermoconductivity $\kappa_{zz}$ also exhibits the $B$-scaling behavior.
{In Fig.~\ref{fig5}, we plot $\kappa_{zz}(B)/\kappa_{zz}(0)$ as a function of $B$ for $\rm{MnBi}_{2}\rm{Te}_{4}$ flakes 
below the Néel temperature~\cite{deng2020sci,otrokov2019nature}},
which exhibits linear dependence of $B$.
Our results show that as long as the system is in the TLMR regime,
other transport coefficients such as the thermoconductivity
also exhibit linear $B$-dependence.

\section{Concluding remarks}\label{con}
In conclusion, we have uncovered a general mechanism
for the TLMR and TLTC
which is induced by the noncentrosymmetric distribution of
the Berry curvature. Such effects are predicted to exist 
in both 2D and 3D systems with nonvanishing Berry curvature.
Our theory provides a satisfactory explanation to the existing 
experiment on the 2D $\rm{MnBi}_2\rm{Te}_4$ flakes~\cite{deng2020science}.
It is also straightforward to explain the same phenomena observed
in the 3D magnetic Weyl semimetal $\rm{Co}_3\rm{Sn}_2\rm{S}_2$ very recently~\cite{moghaddam2022arxiv}.
Actually, Fig.~6 in Ref.~\cite{moghaddam2022arxiv} resembles Figs.~\ref{fig2}(c) and \ref{fig3}(c).

Finally, it is worthwhile to compare our mechanism 
with those in Refs.~\cite{ma2019prb,das19prb,zhang22arxiv}.
Firstly, the LMR in Refs.~\cite{ma2019prb,das19prb} originates from
noncentrosymmetric distribution of the velocity in the reciprocal space while
the TLMR here stems from the noncentrosymmetric Berry curvature. 
Secondly, different from Refs~\cite{ma2019prb,das19prb,zhang22arxiv} that focus 
on the specific Weyl semimetal, our theory offers a general perspective 
that can be applied to a wide range of physical systems.

\section{Acknowledgement}
We thank for the helpful discussion with Ming-Xun Deng and Xing-Yu Liu.
This work was supported by the National Natural Science Foundation of
China under Grant No. 12074172 (W.C.), No. 12222406 (W.C.),
No. 12174182 (D.Y.X), No. 12274203 (B.F.M) and No. 12264019 (W.L.), Fundamental Research
Funds for the Central Universities(W.C.),the startup grant at Nanjing
University (W.C.), the State Key Program for Basic Researches of China
under Grants No. 2021YFA1400403 (D.Y.X.)
and the Excellent Programme at Nanjing University.


%

\end{document}